# Pattern Reification as the Basis for Description-Driven Systems

Florida Estrella[1], Zsolt Kovacs[2], Jean-Marie Le Goff[2], Richard McClatchey[1], Tony Solomonides[1] & Norbert Toth[1]

[1]Centre for Complex Cooperative Systems, UWE, Frenchay, Bristol BS16 1QY UK
Richard.McClatchey@cern.ch
[2]EP Division, CERN, Geneva, Switzerland
Jean-Marie.Le.Goff@cern.ch

**Abstract.** One of the main factors driving object-oriented software development for information systems is the requirement for systems to be tolerant to change. To address this issue in designing systems, this paper proposes a pattern-based, object-oriented, description-driven system (DDS) architecture as an extension to the standard UML four-layer meta-model. A DDS architecture is proposed in which aspects of both static and dynamic systems behavior can be captured via descriptive models and meta-models. The proposed architecture embodies four main elements - firstly, the adoption of a multi-layered meta-modeling architecture and reflective meta-level architecture, secondly the identification of four data modeling relationships that can be made explicit such that they can be modified dynamically, thirdly the identification of five design patterns which have emerged from practice and have proved essential in providing reusable building blocks for data management, and fourthly the encoding of the structural properties of the five design patterns by means of one fundamental pattern, the Graph pattern. A practical example of this philosophy, the CRISTAL project, is used to demonstrate the use of description-driven data objects to handle system evolution.

## 1    Introduction

Many approaches have been proposed to address aspects of design and implementation for modern object-oriented systems. Each has its merits and focuses on concerns such as data modeling, process modeling, state modeling and lifecycle modeling. More or less successful attempts have been made to combine these approaches into modeling languages or methodologies such as OMT [1] and UML [2] but ultimately these approaches lack cohesion since they often represent collections of disparate techniques. Recent reports on their use have led to proposals for enhancements such as pUML [3] and MML [4], which have recognized and begun to address these failings. This paper advocates a design and implementation approach that is systemic (or holistic) in nature, viewing the development of modern object-oriented software from a systems standpoint. The philosophy that has been investigated is based on the systematic capture of the description of systems elements covering multiple views of the system to be designed (including data, process and time views) using familiar techniques. The approach advocated here has been termed *description-driven* and its underlying patterns are the subject of this paper. Essentially the description-driven approach involves identifying and abstracting the crucial elements (such as items, processes, lifecycles, goals, agents and outcomes) in the system under design and creating high-level descriptions of these elements which are stored, dynamically modified and managed separately from their instances.

In the object oriented community well-known design patterns [5] have been named, described and cataloged for reuse by the community as a whole. This approach has enabled us to make use of design patterns that were proven on previous projects and is an example of reuse at the larger grain design level. Our studies have also benefited from the use of frameworks: reusable semi-complete applications that can be specialized to produce custom applications. Frameworks specify reusable architectures for all or part of a system and may include reusable classes, patterns or templates. We note that frameworks focus on the reuse of concrete design algorithms and implementations in a particular programming language, they can be viewed as the reification of families of design patterns and are an important step towards the provision of a truly holistic view of systems design.

Future information systems require more powerful data modeling techniques that are sufficiently expressive to capture a broad class of applications. Evidence suggests that the data model must be object-oriented (OO), since that is the model providing most generality. The data model needs to be an open OO model, thereby coping with different domains having different requirements on the data model [6]. Our studies indicate that use of object meta-models allows systems to have the ability to *model* and *describe* both the static properties of data and their dynamic relationships, to address issues regarding the complexity explosion and the need to cope with evolving requirements as well as the systematic application of software reuse. Making data descriptions (such as information about how data items are organized and related to one another) and system descriptions (such as information about how components are specified and interrelated) run-time accessible allows objects to be composed and managed dynamically.

To be able to describe system and data properties, object meta-modeling makes use of meta-data. The judicious use of meta-data can lead to heterogeneous, extensible and open systems as shown in [7]. Meta-data makes use of an underlying meta-model to describe domains. Research has shown that meta-modeling creates a flexible system offering the following - reusability, complexity handling, version handling, system evolution and interoperability [8]. Promotion of reuse, separation of design and implementation and reification are some further reasons for using meta-models [9]. As such, meta-modeling is a powerful and useful technique in designing domains and developing dynamic systems.

The use of UML, Patterns, Frameworks and OO as design languages and devices clearly eases difficulties inherent in the timely delivery of large complex object-based systems. However, each approach addresses only part of the overall 'design space' and fails to enable a truly holistic view of the design process. In particular they do not easily model the description aspects or meta-information emerging from systems design. In other words, these approaches can locate individual pieces in the overall design puzzle but do not enable the overall puzzle to be viewed. In the next section we look at reflection as the means to open up the design puzzle. This paper then discusses the reification (or explicit materialization) of semantic relationships between objects as the mechanism by which evolving descriptions can be handled and in the following sections identifies a set of meta-objects and their patterns that underpin the reification of semantic relationships. The reified Graph Pattern is isolated as the common thread for these meta-objects and its use in Description-Driven Systems (DDS) is explained in the remainder of this paper with a practical example of a DDS, that of the CRISTAL project under investigation at CERN, Geneva.

## 2    Reflection in Description-Driven Systems

A crucial factor in the creation of flexible information systems dealing with changing requirements is the suitability of the underlying technology to handle the evolution of the system. Exposing the internal system architecture opens up the architecture, consequently allowing application programs to inspect and alter implicit system aspects. These implicit system elements can serve as the basis for changes and extensions to the system. Making these internal structures explicit allows them to be subject to scrutiny and interrogation.

Open architectures can result from a design approach where implicit system aspects (such as system descriptions) are promoted to become (or reified as) explicit first-class meta-objects [10]. Reflective systems are designed to utilize such open architectures. The advantage of reifying system descriptions as objects is that operations can be carried out on them, such as composing and editing, storing and retrieving, organizing and reading. Examples include modifying parts of the implementation strategy such as instance representation, altering language semantics such as relationship behavior and extending the language itself by introducing new data types or new control structures. Meta-objects, as defined here, therefore are self-representations of the system describing how its internal elements can be accessed and manipulated. Since these meta-objects can represent system descriptions, their manipulation can result in a change in the system behavior. As such, reified system descriptions are mechanisms which can lead to dynamically modifiable systems. These self-representations are causally connected to the internal structures they represent; changes to these self-representations immediately affect the underlying system. The ability to dynamically augment, extend and re-define system specifications can result in a considerable improvement in flexibility. This leads to dynamically modifiable systems which can adapt and cope with evolving requirements.

There are a number of OO design techniques which encourage the design and development of reusable objects. In particular design patterns are useful for creating reusable OO designs [5]. Design patterns for structural, behavioral and architectural modeling have been documented and have provided software engineers rules and guidelines which they can immediately (re-)use in software development. In OO programming, the class of a class object is referred to as a meta-class. Meta-objects, therefore, are implemented as meta-classes. Object models used in most class-based programming language are fixed and closed. These object models do not allow the introduction and extension of modeling primitives to cater for specific application needs. The concept of meta-classes is a key design technique in improving the reusability and extensibility of these languages. VODAK [11], ADAM [12] and OMS [13] are some of the next generation DBMSs which have adopted the meta-class approach for tailoring the data model to adapt to evolving specifications.

Reflection figures a significant role in defining a Description-Driven System (DDS) [14]. In our definition a DDS utilizes two modeling abstractions – the model abstraction inherent in multi-layered meta-modeling approach (e.g. OMG's four layer modeling architecture) together with the information abstraction which separates descriptive information (or meta-data) from the data they are describing. DDS make use of meta-objects to store diverse domain-specific system descriptions (such as items, processes, lifecycles, goals, agents and outcomes) which control and manage the life cycles of instances or domain objects. The separation of descriptions from their instances allows them to be specified and managed and to evolve independently and asynchronously. This separation is essential in handling the complexity issues facing many modern computing applications and allows the realization of interoperability, reusability and system evolution as it gives a clear boundary between the application's basic functionalities from its representations and controls. As objects, the reified system descriptions of DDSs can be organized into libraries or frameworks dedicated to the modeling of languages in general, and to customizing its use for specific domains in particular. A practical example of a DDS is presented in Section 6.

This paper describes an investigation of reified design patterns carried out in the context of the CRISTAL project at CERN, Geneva. It shows how the approach of reifying a set of design patterns can be used as the basis of the description-driven architecture for CRISTAL and can provide the capability of system evolution. (The project is not described in detail here. Readers should consult [15], [16], [17] for further detail). The next section establishes how semantic relationships in description-driven systems can be reified using a complete and sufficient set of meta-objects that cater for Aggregation, Generalization, Description and Dependency. In section 4 of this paper the reification of the Graph Pattern is discussed and section 5 investigates the use of this pattern in a three-layer reflective architecture.

## 3      Reifying Semantic Relationships

In response to the demand to treat associations on an equal footing with classes, a number of published papers have suggested the promotion of the relationship construct as a first-class object (reification) [18]. A first-class object is an object which can be created at run-time, can be passed as an actual parameter to methods, can be returned as a result of a function and can be stored in a variable. Reification is used in this paper to promote associations to the same level as classes, thus giving them the same status and features as classes. Consequently, associations become fully-fledged objects in their own right with their own attributes representing their states, and their own methods to alter their behavior. This is achieved by viewing the relationships themselves as patterns.

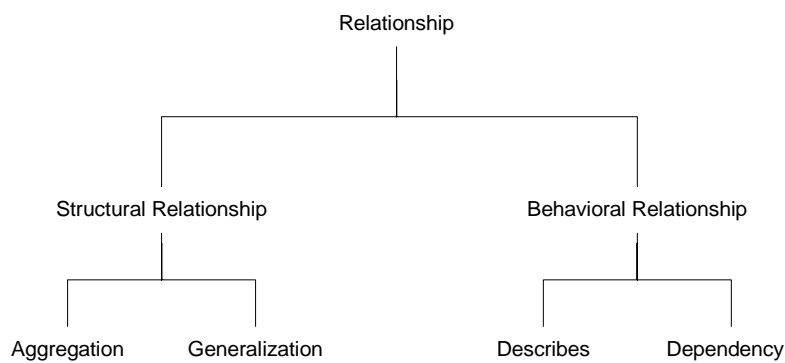

**Figure 1: Relationship classification**

Different types of relationships, representing the many ways interdependent objects are related, can be reified. The proper specification of the types of relationships that exist among objects is essential in managing the relationships and the propagation of operations to the objects they associate. This greatly improves system design and implementation as the burden for handling dependency behavior emerging from relationships is localized to the relationship object. Instead of providing domain-specific solutions to handling domain-related dependencies, the relationship objects handle inter-object communication and domain consistency implicitly and automatically.

Reifying relationships as meta-objects is a fundamental step in the reification of design patterns. The next sections discuss four types of relationships, as shown in Figure 1. The relationship classification is divided into two types - structural relationship and behavioral relationship. A structural relationship is one which deals with the structural or static aspects of a domain. The Aggregation and the Generalization relationships are examples of this type. A behavioral relationship, as the name implies, deals with the behavioral or dynamic aspects of a domain. Two types of behavioral relationships are explored in this paper - the Describes and Dependency relationships.

It is not the object of this paper to give an exhaustive discussion of each of these relationships. Those which are covered are the links which have proved essential in developing the concepts of description-driven systems and these have emerged from a set of five design patterns: the Type Object Pattern [19], the Tree Pattern, the Graph Pattern, the Publisher-Subscriber Pattern and the Mediator Pattern [20]. Interested readers should refer to [21], [22] & [23] for a more complete discussion about the taxonomy of semantic relationships.

### 3.1  The Aggregation Meta-object

Aggregation is a structural relationship between an object *whole* using other objects as its *parts*. The most common example of this type of relationship is the bill-of-materials or parts explosion tree, representing part-whole hierarchies of objects. The familiar Tree Pattern [24] models the Aggregation relationship and the objects it relates. Aggregated objects are very common, and application developers often re-implement the tree semantics to manage part-whole hierarchies. Reifying the Tree pattern provides developers with the Tree pattern meta-object, providing applications with a reusable construct. An essential requirement in the reification of the Tree pattern is the reification of the Aggregation relationship linking the nodes of the tree. For this, aggregation semantics must first be defined.

Typically, operations applied to whole objects are by default propagated to their aggregates. This is a powerful mechanism as it allows the implicit handling of the management of interrelated objects by the objects themselves through the manner in which they are linked together. By reifying the Aggregation relationship, the three aggregation properties of transitivity, anti-symmetry and propagation of operations can be made part of the Aggregation meta-object attributes and can be enforced by the Aggregation meta-object methods. Thus, the state of the Aggregation relationship and the operations related to maintaining the links among the objects it aggregates are localized to the link itself. Operations like copy, delete and move can now be handled implicitly and generically by the domain objects irrespective of domain structure.

Figure 2 illustrates the inclusion of the reified Aggregation relationship in the Tree pattern. In the diagram, the reified Aggregation relationship is called Aggregation, and is the link between the nodes of the tree. The Aggregation meta-object manages and controls the link between the tree nodes, and enforces the propagation of operations from parent nodes to their children. Consequently, operations applied to branch nodes are by default automatically propagated to their compositions.

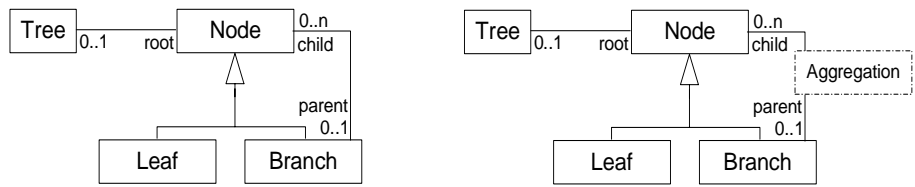

**Figure 2: The Tree Pattern with Reified Aggregation Relationship**

### 3.2 The Generalization Meta-object

Generalization is a structural relationship between a superclass and its subclasses. The semantics of generalization revolve around inheritance, type-checking and reuse, where subclasses inherit the attributes and methods defined by their superclass. The subclasses can alter the inherited features and add their own. This results in a class hierarchy organized according to similarities and differences. Unlike the Aggregation relationship, the generalization semantics are known and implemented by most programming languages, as built-in constructs integrated into the language semantics. This paper advocates extending the programming language semantics by reifying the Generalization relationship as a meta-object. Consequently, programmers can access the generalization relation as an object, giving them the capability of manipulating superclass-subclass pairs at run-time. As a result, application programs can utilize mechanisms for dynamically creating and altering the class hierarchy, which commonly require re-compilation for many languages.

Similar to the Aggregation relationship, generalization exhibits the transitivity property in the implicit propagation of attributes and methods from a superclass to its subclasses. The transitivity property can also be applied to the propagation of versioning between objects related by the Generalization relationship. Normally, a change in the version of the superclass automatically changes the versions of its subclasses. This behavior can be specified as the default behavior of the Generalization meta-object. Figure 3 illustrates the Tree pattern with the Generalization and Aggregation relationships between the tree nodes reified.

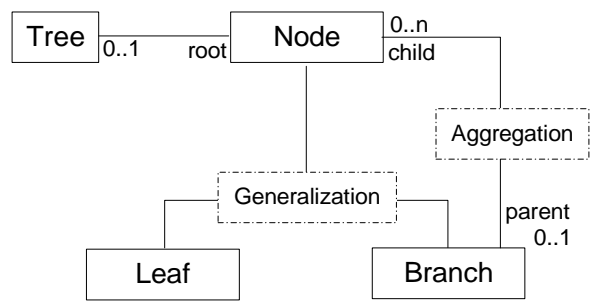

Figure 3: Reification of the Generalization and Aggregation Relationships

**3.3 The Describes Meta-object**

In essence the Type Object pattern [19] has three elements, the object, its type and the Describes relationship, which relates the object to its type. The Type Object pattern illustrates the link between meta-data and data and the Describes relationship that relates the two. Consequently, this pattern links levels of multi-level systems. The upper meta-level meta-objects manage the next lower layer's objects. The meta-data that these meta-objects hold describe the data the lower level objects contain. Consequently, the Type Object pattern is a very useful and powerful tool for run-time specification of domain types.

The reification of the Describes relationship as a meta-object provides a mechanism for explicitly linking object types to objects. This strategy is similar to the approach taken for the Aggregation and Generalization relationships. The Describes meta-object provides developers with an explicit tool to dynamically create and alter domain types, and to modify domain behavior through run-time type-object alteration.

The Describes relationship does not exhibit the transitivity property. This implies that the propagation of some operations is not the default behavior since it cannot be inferred for the objects and their types. For example, versioning a type does not necessarily mean that objects of that type need to be versioned as well. In this particular case, it is the domain which dictates whether the versioning should be propagated or not. Thus, the Describes meta-object should include a mechanism for specifying propagation behavior. Consequently, programmers can either accept the default relationship behavior or override it to implement domain-specific requirements.

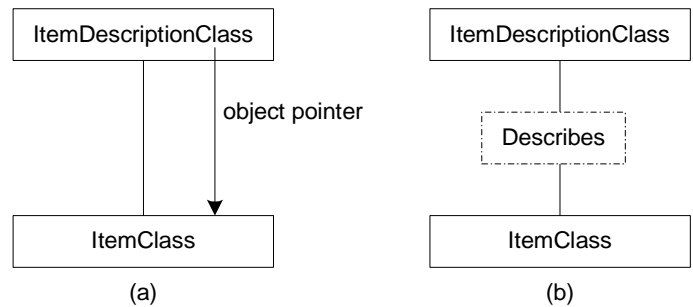

**Figure 4: The Type Object Pattern with Reified Describes Relationship**

Figure 4 illustrates the transformation of the Type Object pattern with the use of a reified Describes relationship. The object pointer (in Figure 4a) is dropped as it is insufficient to represent the semantics of the link relating objects and their types. Instead, the Describes meta-object (in Figure 4b) is used to manage and control the Type Object pattern relationship.

### 3.4 The Dependency Meta-object

The Publisher-Subscriber pattern models the dependency among related objects. To summarize the Publisher-Subscriber pattern, subscribers are automatically informed of any change in the state of its publishers. Thus, the association between the publisher and the subscriber manages and controls the communication and transfer of information between the two. Reifying the Publisher-Subscriber dependency association (hereafter referred to as the Dependency association), these mechanisms can be generically implemented and automatically enforced by the Dependency meta-object itself and taken out of the application code. This represents a significant breakthrough in the simplification of application codes and in the promotion of code reuse.

The reification of the Dependency relationship is significant in that it provides an explicit mechanism for handling change management and consistency control of data. The Dependency meta-object can be applied to base objects, to classes and types, to components of distributed systems and even to meta-objects and meta-classes. This leads to an homogeneous mechanism for handling inter-object dependencies within and between layers of multi-layered architectures.

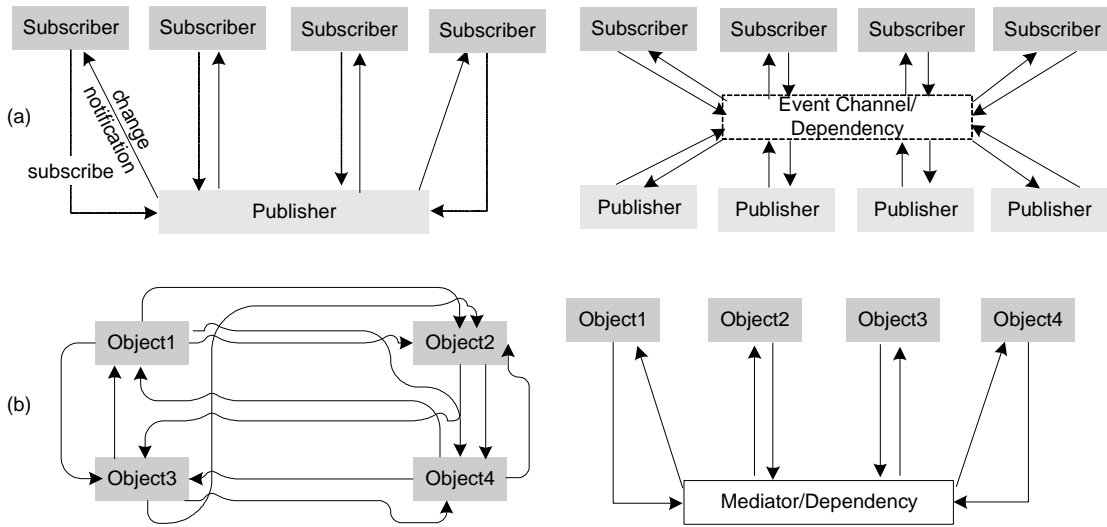

**Figure 5: The Event Channel and the Mediator as Reified Dependency**

The *Event Channel* of the Publisher-Subscriber pattern [25] and the *Mediator* of the Mediator pattern are realizations of the Dependency relationship. The *Event Channel* is an intervening object which captures the implicit invocation protocol between publishers and subscribes. The *Mediator* encapsulates how a set of objects interact by defining a common communication interface. By utilizing the Describes relationship, an explicit mechanism can be used to store and manage inter-object communication protocols. Figure 5 illustrates the use of reified Dependency meta-object in the Publisher-Subscriber pattern (a) and the Mediator pattern (b).

Reifying relationships as meta-objects is a fundamental step in the reification of design patterns. The four relationship meta-objects discussed above manifest the links that exist among the objects participating in the five design patterns listed in the introduction to this section. With the use of reified relationships, these five patterns can be modeled as a single graph, using the Graph pattern. Consequently, the five design patterns can be structurally reified as a Graph pattern, as shown in the next section, with the appropriate relationship meta-object to represent the semantics relating the individual pattern objects.

## 4     The Reified Graph Pattern

The graph and tree data structures are natural models to represent relationships among objects and classes. As the graph model is a generalization of the tree model, the tree semantics are subsumed by the graph model. Consequently, the graph specification is applicable to tree representations. The compositional organization of objects using the Aggregation relationship also forms a graph. Similarly, the class hierarchy using the Generalization relationship creates a graph. These two types of relationships are pervasive in computing, and the use of the Graph pattern to model both semantics provides a reusable solution for managing and controlling data compositions and class hierarchies.

The way dependent objects are organized using the Dependency association also forms a graph. A dependency graph is a representation of how interrelated objects are organized. Dependency graphs are commonly maintained by application programs, and their implementations are often buried in them. The reification of the Dependency meta-object 'objectifies' the dependency graph and creates an explicit Publisher-Subscriber pattern. Consequently, the dependency graph is treated as an object, and can be accessed and manipulated like an object. The same argument applies to the Describes relationship found in the Type Object pattern. The link between objects and their types creates a graph. Reifying the Describes relationship results in the reification of the Type Object pattern. With the reification of the Type Object pattern, the resulting graph object allows the dynamic management of object-type pairs. This capability is essential for environments with unknown or dynamically changing user requirements.

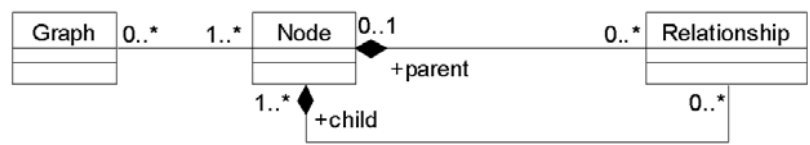

**Figure 6: UML Diagram of the Graph Meta-object**

A UML diagram of the Graph meta-object is shown in Figure 6. The *Node* class represents the entities of the domain objects, classes, data, meta-data or components. The *Relationship* is the reification of the link between the *Nodes*. The aggregated links between the *Node* and the *Relationship* are bidirectional. Two *roles* are defined for the two aggregated associations - that of the *parent*, and that of the *child*. A relationship has at most one parent node, and a parent node can have zero or more relationships. From the child nodes' point of view, a relationship can have at least one child, and a node is a child of zero or more relationships. The parent aggregation, symbolized by the shaded diamond, implies that the lifecycle of the relationship is dependent on the lifecycle of the parent node. The child aggregation behaves analogously.

The use of reflection in making the Graph pattern explicit brings a number of advantages. First of all, it provides a reusable solution to data management. The reified Graph meta-object manages static data using Aggregation and Generalization meta-object relationships, and it makes persistent data dependencies using the Describes and Dependency relationships.

As graph structures are pervasive in many domains, the capture of the graph semantics in a pattern and objectifying them results in a reusable mechanism for system designers and developers. This makes the Graph meta-object a useful guideline applicable to many situations and domains. Another benefit of having a single mechanism to represent compositions and dependencies is its provision for interoperability. With a single framework sitting on top of the persistent data, clients and components can communicate with a single known interface. This greatly simplifies the overall system design and architecture, thus improving system maintainability. Moreover, clients and components can be easily added as long as they comply with the graph interface.

Complexity is likewise catered for since related objects are treated singly and uniformly. Firstly, the semantic grouping of related objects brings transparency to clients' code. Secondly, the data structures provided by the Graph meta-object organize data into atomic units, which can be manipulated as single objects. Objectifying graph relationships allows the implicit and automatic propagation of operations throughout a single grouping. Another benefit in the use of the reified graph model is its reification of the link between meta-data and data. As a consequence, the Graph meta-object not only provides a reusable solution for managing domain-semantic groupings, but can also be reused to manage the links between layers of meta-level architectures.

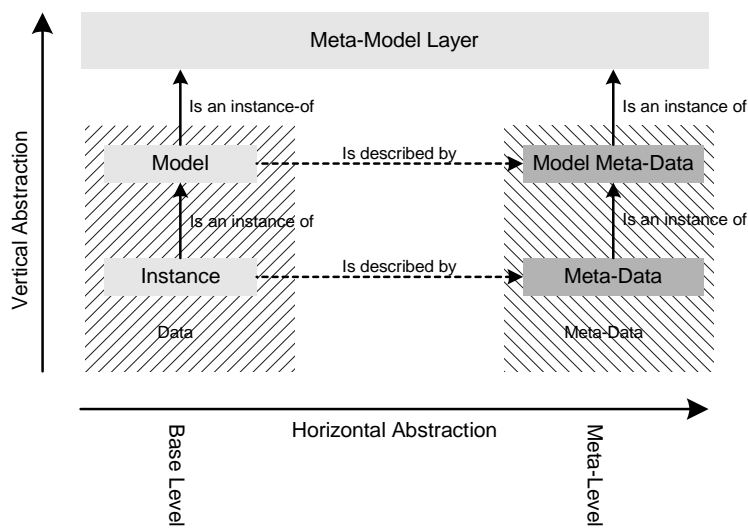

**Figure 7: A Three layer Reflective DDS Architecture**

## 5   Patterns, Relationships and Descriptions

This paper proposes that the reified Graph pattern provides the necessary building block in managing data in any DDS architecture. Figure 7 illustrates a proposed description-driven architecture. The architecture on the left-hand side is typical of layered systems such as the multi-layered architecture specification of the OMG [26]. The relationship between the layers is *Is an Instance-of*. The instance layer contains data which are instances of the domain model in the model layer. Similarly, the model layer is an instance of the meta-model layer. On the right hand side of the diagram is another instance of model abstraction. It shows the increasing abstraction of information from meta-data to model meta-data, where the relationship between the two is also *Is an Instance-of*. These two architectures provide layering and hierarchy based on abstraction of data and information models.

This paper proposes an additional and complimentary view by associating data and meta-data through description (the *Is Described by* relationship). The Type Object pattern makes this possible. The Type Object pattern is a mechanism for relating data to information describing data. The link between meta-data and data using the Describes relationship promotes the dynamic creation and specification of object types. The same argument applies to the model meta-data and its description of the domain model through the Describes relationship. These two horizontal dependencies result in an horizontal meta-level architecture where the upper meta-level describes the lower level. It is the combination of a multi-layered architecture based on the *Is an Instance-of* relationship and that of a meta-level architecture based on the *Is Described by* relationship that results in a DDS architecture.

The reified Graph pattern provides a reusable mechanism for managing and controlling data compositions and dependencies. The graph model defines how domain models are structured. Similarly, the graph model defines how meta-data are instantiated. By reifying the semantic grouping of objects, the Graph meta-object can be reused to hold and manage compositions and dependencies within and between layers of a DDS (see Figure 8).

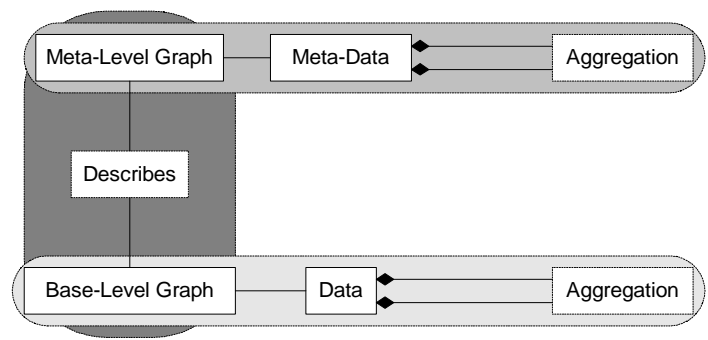

**Figure 8: The Reuse of the Reified Graph Pattern in a DDS**

The meta-level meta data are organized as a meta-level graph. The base-level data are organized as a base-level graph. Relating these two graphs forms another graph, with the nodes related by the Describes relationship. These graphs indicate the reuse of the Graph pattern in modeling relationships in a DDS architecture.

## 6  A Practical Example of a Description-Driven System : The CRISTAL Project

The research which generated this paper has been carried out at the European Centre for Nuclear Research (CERN) [27] based in Geneva, Switzerland. CERN is a scientific research laboratory studying the fundamental laws of matter, exploring what matter is made of, and what forces hold it together. Scientists at CERN build and operate complex accelerators and detectors in an intense research environment in which requirements continuously evolve with time .

The Compact Muon Solenoid (CMS) [28] is a general-purpose experiment that will be constructed from around a million parts and will be produced and assembled in the next decade by specialized centres distributed worldwide. As such, the construction process is very data-intensive, highly distributed and ultimately requires a computer-based system to manage the production and assembly of detector components. In constructing detectors like CMS, scientists require data management systems that are able of cope with complexity, with system evolution over time (primarily as a consequence of changing user requirements and extended development timescales) and with system scalability, distribution and interoperation. No commercial products provide the workflow and product data management capabilities required by CMS.

A research project, entitled CRISTAL (Cooperating Repositories and an Information System for Tracking Assembly Lifecycles [15], [16], [17]) has been initiated to facilitate the management of the engineering data collected at each stage of production of CMS. CRISTAL is a distributed product data and workflow management system which makes use of a commercial database for its repository, a multi-layered architecture for its component abstraction and dynamic object modeling for the design of the objects and components of the system. CRISTAL is based on a DDS architecture using meta-objects. These techniques are critical to handle the complexity of such a data-intensive system and to provide the flexibility to adapt to the changing production scenarios typical of any research production system.

The design of the CRISTAL prototype was dictated by the requirements for adaptability over extended timescales, for system evolution, for interoperability, for complexity handling and for reusability. In adopting a description-driven design approach to address these requirements, the separation of object instances from object descriptions instances was needed. This abstraction resulted in the delivery of a three layer description-driven architecture. The model abstraction (of instance layer, model layer, meta-model layer) has been adopted from the OMG Meta Object

Facility (MOF) specification [29], and the need to provide descriptive information, i.e. meta-data, has been identified to address the issues of adaptability, complexity handling and evolvability.

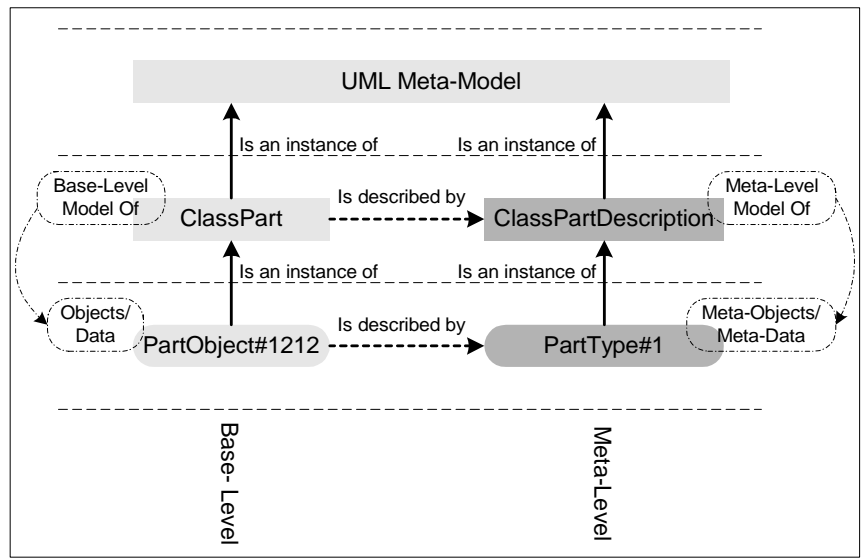

**Figure 9: The CRISTAL DDS Architecture**

Figure 9 illustrates an example of how the DDS is used in the CRISTAL architecture. The CRISTAL model layer is comprised of class specifications for CRISTAL type descriptions (e.g. PartDescription) and class specifications for CRISTAL classes (e.g. Part). The instance layer is comprised of object instances of these classes (e.g. PartType#1 for PartDescription and Part#1212 for Part). The model and instance layer abstraction is based on model abstraction and *Is an instance of* relationship. The abstraction based on meta-data abstraction and *Is described by* relationship leads to two levels - the meta-level and the base-level. The meta-level is comprised of meta-objects and the meta-level model which defines them (e.g. PartDescription is the meta-level model of PartType#1 meta-object). The base-level is comprised of base objects and the base-level model which defines them (e.g. Part is the base-level model of the Part#1212 object).

The CRISTAL meta-object approach reduces system complexity by promoting object reuse and translating complex hierarchies of object instances into (directed acyclic) graphs of object definitions. Meta-objects allow the capture of knowledge (about the object) alongside the object themselves, enriching the model and facilitating self-description and data independence. It is believed that the use of meta-objects provides the flexibility needed to cope with their evolution over the extended timescales of CRISTAL production.

In the CMS experiment, production models change over time. Detector parts of different model versions must be handled over time and coexist with other parts of different model versions.

Separating details of model types from the details of single parts (i.e. the separation of product descriptions from products) allows the model type versions to be specified and managed independently, asynchronously and explicitly from single parts. Moreover, in capturing descriptions separate from their instantiations, system evolution can be catered for while production is underway and therefore provide continuity in the production process and for design changes to be reflected quickly into production.

As the CMS construction is once-off the evolution of descriptions must be catered for. The approach of reifying a set of simple design patterns as the basis of the description-driven architecture for CRISTAL has provided the capability of catering for the evolution of a rapidly changing research data model. In the first two years of operation of CRISTAL it has gathered over 20 Gbytes of data and been able to cope with more than 25 evolutions of its underlying data schema. Detailed discussions of the CRISTAL philosophy can be found in thesis [8] and [20].

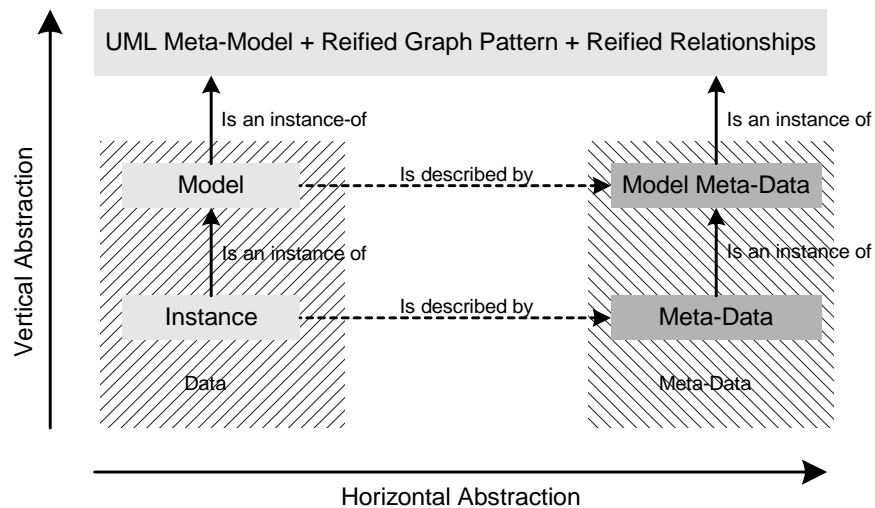

**Figure 10: Extending the UML Meta-model using a Reified Graph Pattern.**

## 7   Related Work and Conclusions

As shown in Figure 10, the reified Graph pattern and the reified relationships enrich the meta-model layer by giving it the capability of creating and managing groups of related objects [30]. The extension of the meta-model layer to include constructs for specifying domain-semantic groupings is the proposition of this paper. The meta-model layer defines concepts used in describing information in lower layers. The core OMG/UML meta-model constructs include *Class, Attribute, Association, Operation* and *Component* meta-objects. The inclusion of the Graph meta-object in the meta-model improves and enhances its modeling capability by providing an explicit mechanism for managing compositions and dependencies throughout the architecture. As a result, the reified Graph pattern provides an explicit homogeneous mechanism for specifying and managing data compositions and dependencies in a DDS architecture.

In comparison, the Active Object Model (AOM) approach of [31] & [32] and recently [33] uses a reflective architecture that can dynamically adapt to new user requirements by storing descriptive information, which can be interpreted at runtime. The primary goal of the AOM is to provide a dynamically evolvable type-safe system. This is achieved by representing traditional class information such as property types, relationships, compositions and object behavior as meta-objects instead of hard-coded information. Class information held by meta-objects are allowed to evolve during run-time, since they are first class objects. As a result, meta-level data represents information about class structure and behaviour. The benefits of such an architecture include run-time changeable object types (classes) and a reflective system where instances can investigate their class structure independent of the OO programming language. Meta-objects in a description-driven system, on the other hand, provide not only structural and behavioral information about instance objects, but act as place holders for all common properties which are required in order to optimally create, instantiate and manipulate the related object instances. Such meta-information may include for example common physical layout data and associated constraints.

Another approach, predicated on 'agile development' rather than reflection, is the contract-oriented coordination method of Andrade and Fiadeiro [34] & [35]. Their fundamental proposition is that under development pressure in the real world, normal methods of adaptation – subclassing and inheritance – impact too heavily on the fundamental design of the system and so must be avoided. Instead, a mechanism for 'coordination' of essentially fixed components is advocated to orchestrate the interactions between these. They take their inspiration, at least in part, from the 'connectors' of parallel architectures. Their common example, rooted in their experience of the banking world, is the multiplicity of *types* of account that a bank may have to provide in order to remain competitive. It is far better, in this view, to alter the 'contract' between an irreducible, *basic* type of account and different clients than to write new account classes and propagate the changed behaviour throughout the system whenever the market dictates a differently nuanced product. Thus the contract at once mediates and orchestrates the relationship between client and account.

A further justification for the 'contract' approach is based on separation of concerns: coordination is seen as a separate requirement from computation. As they put it, "we should be able to superpose regulators (coordination contracts) on given components of a system in order to coordinate their joint behaviour *without having to modify the way these components are*

*implemented*." [36]. The concept of contracts to coordinate components and the concept of reification of semantic links in DDS are design techniques to address (among others) domain complexity explosion and requirements evolution. Contracts can be regarded (in DDS terms) as a semantical mediator link for component coordination.

This paper has shown how reflection can be utilized in reifying design patterns. It shows, for the first time, how reified design patterns provide explicit reusable constructs for managing domain-semantic groupings. These pattern meta-objects are then used as building blocks for describing compositions and dependencies in a three layer reflective architecture - the so-called DDS architecture. The judicious use and application of the concepts of reflection, design patterns and layered models create a dynamically modifiable system which promotes reuse of code and design, which is adaptable to evolving requirements, and which can cope with system complexity.

The design of distributed complex systems should benefit from the approach proposed by this work. Layering and meta-architectures help in the handling of complexity that is inherent in many of today's systems. The need to integrate and inter-operate over distributed information implies the necessity to provide a common framework among these distributed systems. A meta-model can serve as one possible common ground for many systems. Distributed domains can utilize the meta-model primitives in the specification and management of individual domain specifications and in the exchange of distributed data and information. Such an infrastructure brings transparency among distributed elements of the system, as the meta-model interface hides the complexity and domain-specific semantics.

In conclusion, it is interesting to note that the OMG has recently announced the Model Driven Architecture (MDA) [37] as the basis of future systems integration. Such a philosophy is directly equivalent to that expounded in this and earlier papers on the CRISTAL DDS architecture. OMG's goal is to provide reusable, easily integrated, easy to use, scalable and extensible components built around the MDA. While DDS architectures establish those patterns, which are required for exploiting data appearing at different modeling abstraction layers, the MDA approaches integration and interoperability problems by standardizing interoperability specification at each layer (ie standards like XML, CORBA, .NET, J2EE). The MDA integration approach is similar to the Reference Model for Open Distributed Processing (RM-ODP) [38] strategy of interoperating heterogeneous distributed processes using a standard interaction model. In addition, the Common Warehouse Metamodel (CWM) specification [39] has been recently adopted by the OMG. The CWM enables companies better to manage their enterprise data, and makes use of UML, XML and the MOF. The specification provides a common meta-model for warehousing and acts as a standard translation for structured and unstructured data in enterprise repositories, irrespective of proprietary database platforms.

**Acknowledgments**

The authors take this opportunity to acknowledge the support of their home institutes. Nigel Baker, Alain Bazan, Andrew Branson, Peter Brooks, Guy Chevenier, Thierry Le Flour, Sebastien Gaspard, Christoph Koch, Sophie Lieunard, Steve Murray and Gary Mathers are thanked for their assistance in developing the CRISTAL software.